\documentclass[12pt]{article}
\usepackage{epsfig,latexsym}

\def\ofo{ { {}_2 \, \! F_1 }}
\def\kofo{ { {}_1 \, \! F_2 }}

\newcommand{\beq}{\begin{equation}}
\newcommand{\eeq}[1]{\label{#1}\end{equation}}
\newcommand{\bea}{\begin{eqnarray}}
\newcommand{\eea}[1]{\label{#1}\end{eqnarray}}

\input epsf
\input{epsf.sty}
\begin{document}

\font\cmss=cmss10 \font\cmsss=cmss10 at 7pt 

\begin{titlepage}
\hfill NYU-TH/00/09/15

 \vskip .1in \hfill hep-th/0009227

\hfill

\vspace{20pt}

\begin{center}
{\Large \textbf{On the Canonical $c$-Function in 4-d Field Theories 
Possessing Supergravity Duals }}
\end{center}

\vspace{6pt}

\begin{center}
\textsl{ M. Porrati and A.
Starinets } \vspace{20pt}


\textit{Department of Physics, New York University,\\
 4 Washington Place, New York NY 10003 USA}

\end{center}

\vspace{12pt}

\begin{center}
\textbf{Abstract }
\end{center}

\vspace{4pt} {\small \noindent 
 We study monotonicity and other properties of the 
canonical $c$-function
 (defined through the correlator of the energy-momentum tensor)  
in some holographic duals of 4-d  quantum field theories.
The canonical $c$-function and its derivatives are
 related to the $5d$ Green's function of the dual supergravity theory.
While positivity of the canonical $c$-function is obvious, we have not found
a general proof of its monotonicity, even though $c$ is monotonic in the
few explicit examples we examine in this paper.}
\vfill
\end{titlepage}
\eject 
\noindent
Superconformal algebras in 4 dimension possess
two central charges, $c$ and $a$, defined as coefficients 
in the trace anomaly in the background metric $g_{\mu\nu}$:
\begin{equation}
\langle T^{\mu}_{\mu}\rangle_{ren} = \frac{c}{16\pi^2}
 (W_{\mu\nu\rho\sigma})^2 -  \frac{a}{16\pi^2}
 (\tilde{R}_{\mu\nu\rho\sigma})^2 \,,
\label{anomaly}
\end{equation}
where $W_{\mu\nu\rho\sigma}$ is the Weyl tensor and
 $\tilde{R}_{\mu\nu\rho\sigma}$ is the dual of the curvature 
tensor~\cite{afgj}.
Outside the critical points, (\ref{anomaly}) contains  additional terms
proportional to beta-function, with $c$ and $a$ promoted to the
{\it central functions} $c(x)$, $a(x)$, 
since they acquire dependence on the running coupling $g(x)$.
One particularly interesting subclass of
these theories, characterized by $c = a$, has properties closely
resembling those of the two-dimensional CFTs \cite{a1}. All theories 
admitting a dual description in terms of supergravity fall in this class.
The converse has also been conjectured to be true~\cite{a22}.
In this case one can use $5d$ asymptotically
AdS backgrounds to study RG flows between UV 
 and IR conformal fixed points
induced by relevant deformations of the $4d$ SCFT. Just as in 
two dimensions, the flows are 
{\it irreversible}, having $c_{UV}-c_{IR}\geq 0$. 
On the gravity side of the 
AdS/CFT, there exists a natural candidate for the Zamolodchikov-type
$c$-function monotonically decreasing along the flow as a function of 
a scale - courtesy of Einstein's equations of motion\cite{gppz1,freed1}.
 This {\it holographic}
$c$-function, $c_H$, coincides with the canonical $c$-function
defined by the equation~\cite{Cappelli,a3} 
\begin{equation}
\langle T_{\mu\nu} (x) T_{\rho\sigma}(0)\rangle = -{\frac{1}{48\pi^4}}{\Pi}%
^{(2)}_{\mu\nu\rho\sigma} \left[{\frac{c(x)}{x^4}}\right] +
\pi_{\mu\nu}\pi_{\rho\sigma}\left[{\frac{f(x)}{x^4}}\right],
\label{duepunti}
\end{equation}
where 
$\pi_{\mu\nu}=\partial_\mu\partial_\nu -\eta_{\mu\nu}\partial^2 $, and 
${\Pi}^{(2)}_{\mu\nu\rho\sigma}=2\pi_{\mu\nu}\pi_{\rho\sigma} -3 
(\pi_{\mu\rho}\pi_{\nu\sigma}+\pi_{\mu\sigma}\pi_{\nu\rho})$,
at the critical UV and IR points, where both are equal to the 
central charges $c_{UV}$ and $c_{IR}$, respectively \cite{Cappelli,a3}.
Outside the critical points, the $c$ -function is not unique;
in particular, the holographic $c$-function $c_H$ does not (and need not to)
coincide with the canonical one. There is also no {\it a priori} reason for
the canonical $c$-function to be monotonic, although its values
at the UV and IR critical points  still satisfy inequality
$c_{UV}\geq c_{IR}$. In fact, it is known that outside the $c=a$ corridor
the canonical $c$-function is {\it not} monotonic \cite{a1}.
However, since theories that possess supergravity duals 
are distinguished by having $c=a$ 
(to the leading order in $N$), and 
the supergravity description of RG flows naturally exhibits
monotonicity, one may think that the canonical $c$-function 
somehow also inherits the monotonicity from the supergravity data.
This is indeed so in the models studied so far, but we still do not know
 whether or not this is true in general.

In this letter we study properties of the canonical $c$-function defined by
the Eq.(\ref{duepunti}) by relating it and its derivatives 
to the $5d$ Green's function of the Klein - 
Gordon equation  in the asymptotically AdS background.
We shall use coordinate system in which $5d$ metric has the form
\begin{equation}
ds^2=dy^2 + e^{2\phi (y)} dx^{\mu}dx_{\mu}\,.
\label{metric}
\end{equation}
In the supergravity description of the
dual $4d$ RG flows the metric (\ref{metric}) is typically  
characterized by the following properties:

{\em i)} It is asymptotically AdS, i.e.
$\phi (y) = y/R_{UV} + O(e^{-2y/R_{UV}})$ for $y\rightarrow\infty$,
where  $R_{UV}$ is the ``radius'' of
the AdS space at $y\rightarrow \infty$. We shall set $R_{UV} =1$ henceforth.

{\em ii)} It has 
 an IR  singularity
at some finite\footnote{ In the conformal case instead of a 
singularity there exists a horizon located at $y=\infty$.}
 value of $y$ which we can set to zero without loss of generality.

{\em iii)} If $\phi (y)$ is the function describing the background of the 
supergravity dual to a certain gauge theory with $4d$ coordinate $x$,
then $\phi (y+ \lambda ) - \lambda$ is the background corresponding to
the rescaled $4d$ coordinate $e^{-\lambda}x$ (with the IR singularity 
now located at $y= -\lambda$,  $\phi (y)\rightarrow -\infty $
 for $y\rightarrow - \lambda$). From points $i)$ and $iv)$ we also have:

{\em iv)} The derivative of $\phi$ with respect to $y$ is 
a monotonic function, $\phi_{yy}''\leq 0$, $\phi_{yy}''= 0$
at the critical points.

{\em v)} $d\phi (y)/dy \geq 1$.

{\em vi)}  $d\phi (y)/d \lambda  = d\phi (y)/d y  - 1$.

The transverse traceless part of the energy-momentum tensor,
\begin{equation}
{\cal T}_{\mu\nu} =  T_{\mu\nu} + \frac{1}{3} \frac{1}{\Box} 
\pi_{\mu\nu} T\,,
\end{equation}
where $T=T^{\mu}_{\mu}$,  obeys
\begin{equation}
\langle {\cal T}_{\mu\nu}(x)  {\cal T}_{\rho\sigma}(0) \rangle
 \,=\, - \frac{1}{48 \pi^4} \Pi_{\mu\nu\rho\sigma}^{(2)}
 \left( \frac{c(x)}{x^4} \right)\,.
\end{equation}
With the Euclidean 4-momentum $k$ oriented along $z$-axis we find
\begin{equation}
c(|x|) = |x|^4 \int e^{-i k x} \frac{{\cal F}(k^2)}{k^4} d^4k = 4\pi^2 |x|^3 
 \int\limits_{0}^{\infty} \frac{J_1(k|x|)}{k^2}
{\cal F}(k^2)dk\,,
\end{equation}
where ${\cal F}(k^2) = \langle {\cal T}_{xy}(k)  {\cal T}_{xy}(0) \rangle $ is 
 the momentum space
 two-point function of a minimally coupled massless scalar in
 an asymptotically $AdS$ background. The function  ${\cal F}(k^2)$ can be
 computed following the standard procedure \cite{gkp}, if the 
 solution
of the free wave equation in the asymptotically $AdS$ background is known.
For pure $AdS_5$ we have ${\cal F}(k^2)=-N^2 k^4\log{k^2}/64\pi^2$,
and therefore $G(x)=3N^2/\pi^4 x^8$ and $c(x) = N^2/4$
(in the normalization of~\cite{gkp}).
By dimensional analysis, the canonical central function
$c(x)$ depends on $\lambda$ through the combination $xe^{-\lambda}$.
Since  $\dot{c}(x,\lambda )\equiv x\partial_x c(x,\lambda ) = 
- d c(x,\lambda)/d\lambda$, the desired monotonicity property of $c(x)$,
 $\dot{c}(x,\lambda )\leq 0$,
 is
translated into the monotonicity of $c(x,\lambda )$ with respect 
to $\lambda$, 
$\partial_{\lambda} c(x,\lambda ) \geq 0$.
We now try to establish this property by 
 looking at the $5d$ Green's function $G(x,y|x',y')$. Due to the 
$4d$ translation invariance, $G(x,y|x',y')$ depends on $|x-x'|$; 
we put $x'=0$ and use $x$ instead of $|x|$ 
to simplify notations. The function $G(x,y|0,y')$ obeys 
\beq
-\partial_y e^{4\phi (y)}\partial_y G 
- \Box e^{2\phi (y)} G = \delta^{(4)}(x) \delta (y-y')\,,
\eeq{6}
from which we deduce
\beq
e^{4\phi (y)}\partial_y G (x,y|0,y')\mid_{ren}
= - \Box^2 \int\limits_{-\lambda }^{y}dy_1 e^{2\phi (y_1)}
\int\limits_{y_1}^{\infty}dy_2 e^{-4\phi (y_2)}
\int\limits_{-\lambda }^{y_2}dy_3 e^{2\phi (y_3)} G(x,y_3|0,y')\,,
\eeq{7}
where the symbol $|_{ren}$ means that terms proportional to
 $\delta^{(4)}(x)$ are discarded.
Since
\begin{equation}
c(x,\lambda )=\kappa x^4 \lim_{y\rightarrow\infty} 
\lim_{y'\rightarrow\infty}
e^{4\phi (y)}e^{4\phi (y')}
\partial_y\partial_{y'}\Box^{-2} G (x,y|0,y')\mid_{ren}\,,
\label{8}
\end{equation}
where $\kappa = (2\pi)^4 N^2/32\pi^2$ is the normalization constant,
we obtain 
\begin{equation}
c(x,\lambda )= - \kappa 
x^4 \lim_{y\rightarrow\infty} \lim_{y'\rightarrow\infty}
 e^{4\phi (y')}  \partial_{y'}\int\limits_{-\lambda }^{y}dy_1
\int\limits_{y_1}^{\infty}dy_2
\int\limits_{-\lambda }^{y_2}dy_3 F(x,y_1,y_2,y_3,y',\lambda )\,,
\label{9}
\end{equation}
\begin{equation}
x\partial_x c(x,\lambda ) = -\frac{d}{d\lambda }c(x,\lambda ) 
=   \kappa 
x^4 \lim_{y\rightarrow\infty} \lim_{y'\rightarrow\infty}
 e^{4\phi (y')}  \partial_{y'}\int\limits_{-\lambda }^{y}dy_1
\int\limits_{y_1}^{\infty}dy_2
\int\limits_{-\lambda }^{y_2}dy_3 \; F'_{\lambda}\,,
\label{9a}
\end{equation}
where
\begin{equation}
F(x,y_1,y_2,y_3,y',\lambda ) = e^{2\phi (y_1) - 4\phi (y_2 ) +2\phi (y_3)}
 G(x,y_3|0,y')\,.
\end{equation}
The above expression can also be written as
\begin{eqnarray}
x\partial_x c(x,\lambda ) &=&  
  \kappa x^4 \lim_{y\rightarrow\infty} \lim_{y'\rightarrow\infty}
 e^{4\phi (y')}  \partial_{y'}\int\limits_{-\lambda }^{y}dy_1 e^{2\phi (y_1)}
\int\limits_{y_1}^{\infty}dy_2 e^{-4\phi (y_2)}
\int\limits_{-\lambda }^{y_2}dy_3 e^{2\phi (y_3)} \nonumber \\
&\,& \left[\frac{d G(x,y_3|0,y')}{d \lambda }+
2 G(x,y_3|0,y') (\phi ' (y_1) - 2\phi ' (y_2) + \phi ' (y_3))\right]\,,
\label{eq10}
\end{eqnarray}
where $\phi ' (y)$ denotes the derivative of $\phi$ with respect to $\lambda$.
We would like to see that $dc(x,\lambda )/d\lambda \geq 0$.
Since for $y'\rightarrow \infty$ $G(x,y_3|0,y')\sim e^{-4y'}$,
the sign of $dc(x,\lambda )/d\lambda $ is the same as the sign of
$d G(x,y_3|0,y')/d \lambda  +
2 G(x,y_3|0,y') (\phi ' (y_1) - 2\phi ' (y_2) + \phi ' (y_3))$.
We have also $y_1\leq y_2$, $y_3\leq y_2$
and therefore $\phi ' (y_1) - 2\phi ' (y_2) + \phi ' (y_3) \geq 0$ 
because of properties $v)$ and $vi)$ of $\phi (y)$ itemized above.
Since $G(x,y_3|0,y') > 0$, the second term in the square 
bracket in Eq.(\ref{eq10}) is nonnegative.

We shall investigate now the monotonicity of $G(x,y|0,y')$
with respect to the parameter $\lambda$.
The derivative $\partial_{\lambda} G(x,y|0,y')$ obeys
\beq
\left( -\partial_y e^{4\phi (y)}\partial_y - e^{2\phi (y)}\right)
 \partial_{\lambda} G(x,y|0,y') = R(x,y,y')\,,
\eeq{13}
where 
\beq
R(x,y,y')= 4 \partial_y \phi '(y) e^{4\phi}\partial_y G + 2\phi '(y)
  \partial_y (e^{4\phi}\partial_y G) - 2 \phi '(y) \delta^{(4)}(x)\delta(y-y')
\,.
\eeq{14}
Then
\beq
\partial_{\lambda} G(x,y|0,y') =
 \int\limits_{-\lambda}^{\infty}d\tilde{y}
 \int d\tilde{x} G(x,y|\tilde{x},\tilde{y})R(\tilde{x},\tilde{y}, y')\,.
\eeq{15}
To prove monotonicity of $G(x,y|0,y')$ with respect to 
$\lambda$ we have to show that $R(x,y,y')\mid_{ren}$ given by 
\beq
R(x,y,y')\mid_{ren}  = 4 \partial_y \phi '(y) 
e^{4\phi}\partial_y G + 2\phi '(y)
  \partial_y ( e^{4\phi}\partial_y G)
\,
\eeq{16}
(the term with delta-functions is irrelevant since upon integration 
it produces $-2\phi '(y')$   $G(x,y|0,y')\rightarrow 0$ for
 $y'\rightarrow \infty$ and $x>0$) is nonnegative.
Even though this is certainly true for large values of $y$,
it appears to be {\it false} in general (see 
the counterexample below).\footnote{
 If $G$ is represented by $G=G_0 e^{-4y} + G_1 e^{-6y}+\dots$,
where $G_0$ is positive by unitarity, then $e^{4\phi (y)}
\partial_y G \rightarrow -4G_0 <0$ for large $y$, whereas the term $2\phi '(y)
  \partial_y (e^{4\phi}\partial_y G)$ is suppressed 
by the additional power of $e^{-2y}$ (note that $\phi '(y)$ and 
$\partial_y \phi '(y)$ have the same order of magnitude $\sim e^{-2y}$). 
Since $\partial_y \phi '(y) \leq 0$, 
$R(x,y,y')\mid_{ren}=$  $-16 e^{-4y'} \partial^2_{yy} \phi (y) G_0(x)
[1 + O(e^{-2y},e^{-2y'})]\geq 0$ and therefore $dG/d\lambda\geq 0$.
}

Indeed, if the property $dG(x,y|0,y')/d\lambda \geq 0$ would hold in general,
it would imply  $dG(x)/d\lambda \geq 0$, where $G(x)$ is
the $4d$ two-point function (the zeroth-order term in the expansion of
$G(x,y|0,y')$ in powers of $e^{-4y}$, $e^{-4y'}$).
The behavior of $G(x)$ with respect to $\lambda$ is related to the
behavior of the scale-invariant function 
$x^8G(x)$ with respect to $x$,
since $dG/d\lambda$  $= - x\partial_x (x^8G(x))/x^8$. 
We shall demonstrate that in one of the three known explicit supergravity
realizations of the dual RG flows, the function $x^8G(x)$ is not
monotonic in $x$, and therefore  $dG(x,y|0,y')/d\lambda \geq 0$ can
not be true in general.
Intriguingly, in spite of this negative result, $c(x)$ turns out to be 
monotonic in all explicit examples described below.

\begin{figure}[h]
\begin{center}
\epsffile{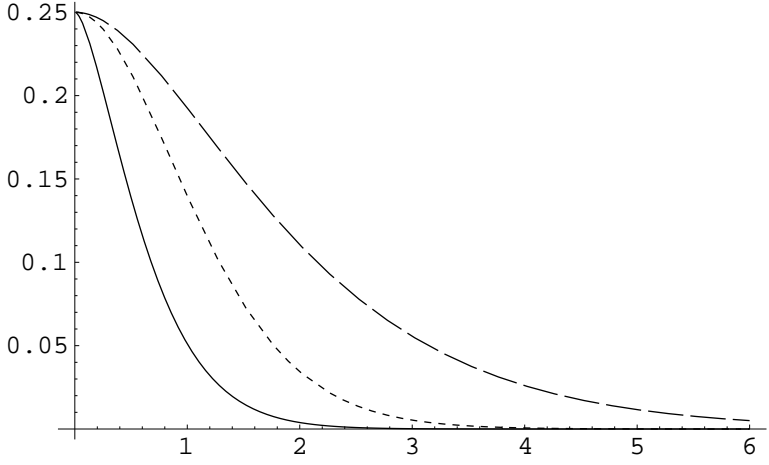}
\end{center}
\caption{Function $c(x)/N^2$ for the flow to ${\cal N}=1$ SYM (solid line),
${\cal N}=4$ Coulomb branch, discrete spectrum, $n=4$ case (dotted line),
 ${\cal N}=4$ Coulomb branch, continuous spectrum with a mass gap, $n==2$ case
 (dashed line).}
\label{apr1}
\end{figure}

\subsection*{ Flow to ${\cal N}=1$ SYM}

\begin{figure}[p]
\vspace*{-0.2cm} \hspace*{-0cm}
\begin{center}
\vspace*{0cm} \hspace*{-0cm}
\epsfxsize=0.5\textwidth
\leavevmode\epsffile{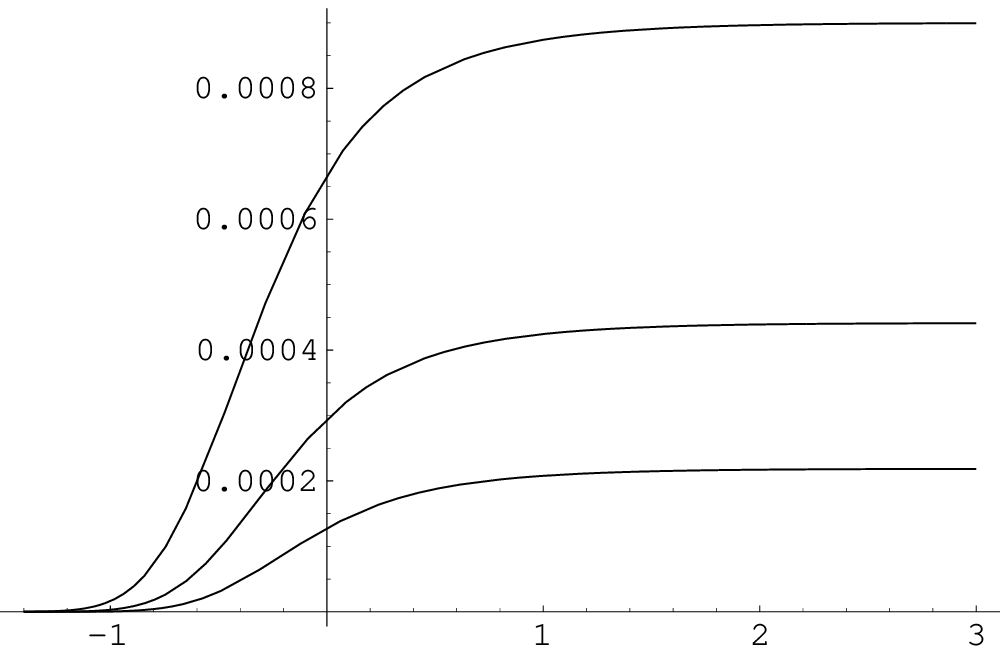} \\
\caption{${\cal N}=1$ flow: five-dimensional Green's function 
$G(x,y=1)$ versus $\lambda$ for (from top to bottom) $x=1.41$,
$x=1.55$, $x=1.7$ }
\label{apr112}
\vspace*{0.7cm}
\epsfxsize=0.5\textwidth
\leavevmode\epsffile{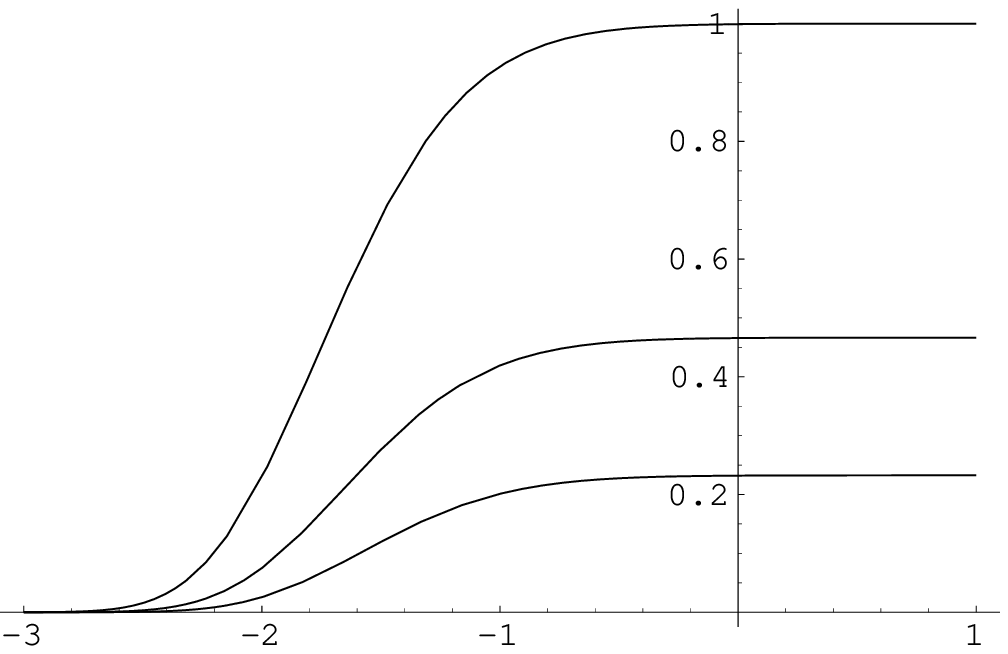} \\
\caption{${\cal N}=4$ flow, $n=2$ (continuous spectrum):
four-dimensional 
Green's function $G(x)$  
versus $\lambda$ for (from top to bottom) $x=1$,
$x=1.1$, $x=1.2$ }
\label{apr114}
\vspace*{0.7cm}
\epsfxsize=0.5\textwidth
\leavevmode\epsffile{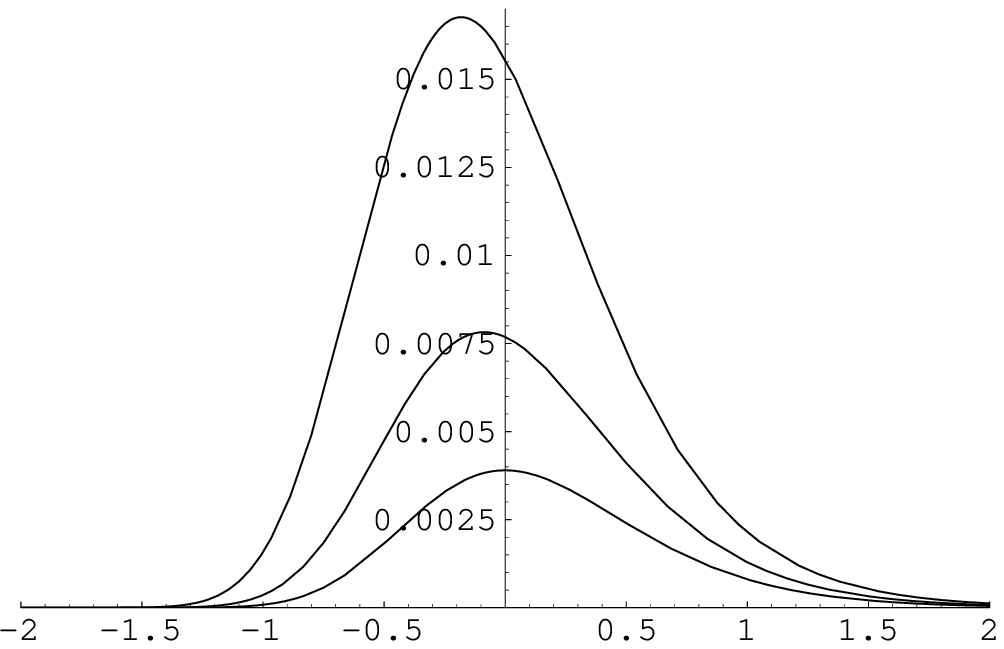}
\caption{${\cal N}=4$ flow, $n=4$ (discrete spectrum):
four-dimensional 
Green's function $G(x)$ 
versus $\lambda$ for (from top to bottom) $x=1$,
$x=1.1$, $x=1.2$ }
\label{apr113}
\end{center}
\end{figure}

In our first example, we calculate the $5d$ Green's function
and the central function $c(x,\lambda)$
for the flow to the  ${\cal N}=1$ SYM studied in~\cite{gppz3}.
The metric is of the form (\ref{metric}),
where $\phi (y)$ is given by
\begin{equation}
\phi (y) = \frac{1}{2} \left( y +
 \log \left[ 2 \sinh (y+ \lambda )\right] -\lambda \right)\,.
\label{2}
\end{equation}
The function $\phi (y)$ has properties {\it i) - vi)} listed in the
beginning of the paper.
The solutions of the Klein-Gordon equation in the background given by (\ref{2})
are
\begin{equation}
\phi_1 (y) = \frac{(1-A^2)\pi A}{\sin{\pi A}}
 \ofo \left( A, - A; 2; 1-e^{-2y-2\lambda }\right)\,,
\end{equation}
\begin{equation}
\phi_2 (y) = - e^{-4y} \ofo \left( 2+A, 2- A; 3; e^{-2y-2\lambda }\right)\,,
\end{equation}
where $A=ike^{\lambda}/2$.
The Green's function obeying Dirichlet condition at the boundary ($y=\infty$) 
is constructed in a standard way, and in the region $-\lambda \leq y \leq y'$
with $y'\rightarrow \infty$ is represented by a series
\begin{equation}
G(k^2,y,y') = e^{-4y'} 
\frac{(1-A^2)\pi A}{\sin{\pi A}}
 \ofo \left( A, - A; 2; 1-e^{-2y-2\lambda }\right) + O(e^{-6y'})\,.
\label{Green_k}
\end{equation}
Then the $4d$ two-point function, ${\cal F}(k^2)$, can be
obtained as
\begin{equation}
{\cal F}(k^2) = \frac{N^2}{32\pi^2}
\lim_{y\rightarrow\infty} \lim_{y'\rightarrow\infty}
 e^{4(\phi (y')+\phi (y))} \partial_{y} \partial_{y'}
G(k^2,y,y')\,.
\end{equation}
In the coordinate space (\ref{Green_k}) becomes 
 $G(x,y|0,y') = e^{-4y'}G(x,y) + O(e^{-6y'})$, where
\beq
G(x,y) = \sum\limits_{n=0}^{\infty} G_n(x,y)\,,
\eeq{11}
\beq
G_n(x,y) = - \frac{N^2 e^{-7\lambda }}{\pi^4 x}
 \left( \frac{e^{-2y}}{e^{2\lambda}}\right)^n
\frac{2e^{-4y}}{n!(n+2)!}\sum\limits_{l=n+2}^{\infty} l^4(1^2 - l^2)
\cdots \left( (n+1)^2 -l^2\right) K_1(2 l x e^{-\lambda}).
\eeq{12}
The first term,  $G_0(x,y)=e^{-4y} G(x)$,
 gives the $4d$ two-point function:
\begin{equation}
G(x) = \frac{N^2e^{-7\lambda} }{\pi^4 x} 
\sum\limits_{n=2}^{\infty}n^4(n^2-1) K_1 (2nxe^{-\lambda}) = 
\frac{ N^2e^{-7\lambda}}{\pi^4 x}  
\int\limits_{0}^{\infty}\frac{f(xe^{-7\lambda},t)
\cosh{t}}{(e^{2x\exp{-\lambda}\cosh{t}}-1)^7} dt\,,
\end{equation}
where $f(u,t) = 48 e^{4u\cosh{t}} (1 +  e^{6u\cosh{t}})+ 312  e^{6u\cosh{t}}
 (1 +  e^{2u\cosh{t}}) >0$. 
The $5d$ Green's function is monotonic with respect to the parameter
 $\lambda$ (see Fig.(\ref{apr112})) which according to
the discussion in the preceding section implies monotonicity of the
canonical central function.
The explicit expression for $c(x)$ was obtained in \cite{a_note}:
\begin{equation}
c(x) = N^2  x^3e^{-3\lambda}
 \sum\limits_{n=2}^{\infty}(n^2-1) K_1 (2nxe^{-\lambda}) = 
N^2 x^3 e^{-3\lambda}
\int\limits_{0}^{\infty} 
\frac{3 e^{2x\exp{-\lambda}
\cosh{t}}-1}{(e^{2 x\exp{-\lambda} \cosh{t}}-1)^3}\cosh{t} dt\,.
\label{c_gubser_ne_gubser}
\end{equation}
Using Mellin summation technique\footnote{Mellin summation
(rather than the Euler-Maclaurin formula)
is convenient in this case since the function of $n$ multiplying
the modified Bessel function in (\ref{c_gubser_ne_gubser})
 is a polynomial.
 This is not the case in our third example, where it is 
the Euler-Maclaurin formula which allows to obtain an explicit expression
whilst  the Mellin summation is useless.}
 \cite{haber} we obtain the 
following representation for $c(x)$
\begin{eqnarray}
c(x) &=&
\frac{N^2}{4} - \frac{N^2 x^2 e^{-2\lambda}}{24} + 
\frac{N^2x^2e^{-2\lambda}}{2}\log{ (xe^{-\lambda}})
\nonumber \\ &-&N^2x^4e^{-4\lambda} 
\sum\limits_{n=0}^{\infty}\frac{x^{2n}e^{-2n\lambda}}{n!(n+1)!}
 \left[ \zeta ' (-2n-3) -  \zeta ' (-2n-1) \right] 
\label{mellin_ne_gubser} \\
&-& \frac{N^2x^4e^{-4\lambda}}{2}
 \sum\limits_{n=0}^{\infty}\frac{x^{2n}e^{-2n\lambda}}{n!(n+1)!}\left(
\frac{B_{2n+2}}{n+1} - \frac{B_{2n+4}}{n+2}\right)
\left[ \psi (n+1) + \frac{1}{2(n+1)} - \log{ (xe^{-\lambda}}) \right],
\nonumber 
\end{eqnarray}
where $\zeta '(s)$ is the derivative of the Riemann zeta-function,
$B_{2n}$ are Bernoulli numbers. The central function 
is positive definite, monotonic (Fig. (\ref{apr1})),
 and has the correct UV and IR
 limits: for small $x$   $c(x)\rightarrow c_{UV}=N^2/4$ and  
for $x\rightarrow \infty$
 $c(x)\sim x^{5/2}e^{-4x}\rightarrow c_{IR}=0$.
 It is also non-analytic in $x$
as the expansion (\ref{mellin_ne_gubser}) shows.

\subsection*{Coulomb branch of ${\cal N} = 4$ SYM}
The metric corresponding to the supergravity description of the 
$n$-dimensional ($1\leq n\leq 5$) transverse distribution of the $D3$ branes
(associated with the Coulomb branch of the  ${\cal N} = 4$ SYM with
 $SO(n)\times SO(6-n)$ local gauge symmetry)
was obtained in \cite{Gubser}. For $n=2,4$ the $s$-wave 
Klein  - Gordon equation for a
minimally coupled massless scalar,
\begin{equation}
\frac{1}{r}\partial_r 
\left[ r^5 \left( 1 \pm \frac{\ell^2}{r^2}\right)\partial_r \xi_k (r)\right]
- k^2 L^4 \xi_k (r) = 0\,,
\label{gubser_equation}
\end{equation}
where $\pm$ corresponds to $n=2,4$, respectively,  
$L$ is the radius of the asymptotic $AdS_5$ and $\ell$ characterizes
the size of the branes distribution, can be solved exactly.
We consider the $n=2$ and the $n=4$ cases separately.

\subsubsection*{Two-dimensional distribution of $D3$ branes}
The solutions of (\ref{gubser_equation}) are 
\begin{equation}
\xi_k^{(1)} (r) = v^{a_+} \ofo \left(a_+,a_+;2a_+ +2;v\right)\,,
\end{equation}
\begin{equation}
\xi_k^{(2)} (r) = v^{a_-} \ofo \left(a_-,a_- ; 2a_- +2;v\right)\,,
\end{equation}
where $a_{\pm}=-1/2\pm \sqrt{1+e^{2\lambda} k^2}$, $e^{\lambda} = L^2/\ell$,
$v=1/(1+\ell^2/r^2)$. The first solution is regular at the location
of the branes, and is used to construct the two-point function \cite{Gubser}:
\begin{equation}
{\cal F}(k^2)= - \frac{N^2 k^4}{32 \pi^2} \Psi \left(\frac{1}{2} +
\frac{1}{2}\sqrt{1+ e^{2\lambda} k^2}\right)\,.
\end{equation}
Then the  central function is given by
\begin{equation}
c(x) = - \frac{N^2 x^3}{8} \int\limits_{0}^{\infty} k^2 J_1 (kx) \Psi
 \left( \frac{1}{2} +\frac{1}{2}\sqrt{1 + e^{2\lambda} k^2}\right) dk\,.
\end{equation}
Using the Binet expression \cite{bateman},
\begin{equation} 
\Psi (z) = \log{z} - 1/2z -
 2\int\limits_{0}^{\infty} (t^2+z^2)^{-1} (e^{2\pi t} -1)^{-1} t dt\,,
\end{equation}
and the result\footnote{ This integral can be obtained by taking a derivative
with respect to $c$ of the expression 
$$
\int\limits_{0}^{\infty} x \log{ \left( 1+\frac{z^2}{x^2}\right) }
 J_0 (cx) dx = 
\frac{2}{c^2} - \frac{2z}{c} K_1(cz) \,,
$$
given in  2.12.28.9 of \cite{prudnikov}. }
\begin{equation}
\int\limits_{0}^{\infty} x^2 \log{ \left( 1+\frac{z^2}{x^2}\right) }
 J_1 (cx) dx = 
\frac{4}{c^3} - \frac{2z^2}{c} K_2(cz) \,,
\end{equation}
one can obtain the following spectral representation for $c(x)$:
\begin{equation}
c(x) = \frac{N^2 x^4e^{-4\lambda}}{4}
 \int\limits_{1}^{\infty} \rho(\mu^2) \Delta_E
 (e^{-\lambda}x,\mu^2)d\mu^2\,, 
\end{equation}
where
$$
\rho (\mu^2) = \pi^2 \left( 1 - \frac{2}{e^{\pi \sqrt{\mu^2-1}}-1} +
 \frac{4}{e^{2\pi \sqrt{\mu^2-1}}-1}\right)
$$
and $\Delta_E (x,m^2)=\sqrt{m^2x^2} K_1(\sqrt{m^2x^2})/4\pi^2x^2$
 is a four-dimensional free Euclidean propagator.
For small $x$ we get
\begin{equation}
c(x) = \frac{N^2}{4} - \frac{7 N^2 x^2 e^{-2\lambda}}{96}
 + O\left( x^4, x^4\log{x}\right)\,.
\label{c_gubser_1_small_x}
\end{equation}
Using this expansion, we can calculate the anomalous dimension 
of the operator inducing the flow 
 in the UV limit,
\begin{equation}
h_{UV}=-\lim_{UV}\frac{\ddot{c}}{2\dot{c}} = -1\,.
\end{equation}
The coordinate representation of the scale-invariant 
function $x^8G(x)$ is given by
\begin{eqnarray}
x^8 G (x) &=& \frac{3N^2}{\pi^4 } \Biggl( 1 + x K_1 (x) \frac{(x^2+8)^2}{64} +
 x^2 K_0(x) \frac{x^2+8}{16} \nonumber \\
& + & \frac{x^6}{384} K_0(x) -
x \left[ K_0(x)I_1(x) + K_1(x) I_0(x) \right] \nonumber \\ &+& 
 \frac{x^8}{192}\int\limits_{1}^{\infty}
\frac{\mu^4}{e^{2\pi \sqrt{\mu^2-1}}-1}\frac{\mu K_1(\mu x)}{x}d\mu^2
-\frac{x^8}{384}\int\limits_{1}^{\infty}
\frac{\mu^4}{e^{\pi \sqrt{\mu^2-1}}-1}\frac{\mu K_1(\mu x)}
{x}d\mu^2 \Biggl) \nonumber \,.
\end{eqnarray}
Note that $x^8G(x)$ 
(or $x^8e^{-8\lambda}G(xe^{-\lambda})$ versus $\lambda$)
in this case is monotonic (Fig.(\ref{apr114})) which again implies 
monotonicity of $c(x)$ (Fig.(\ref{apr1})).

\subsubsection*{Four-dimensional distribution of $D3$ branes}

The solutions of (\ref{gubser_equation}) for $n=4$ are
\begin{equation}
\xi_k^{(1)} (r) = (1-u)^2  \ofo \left(a+2,1-a;1;u\right)\,,
\end{equation}
\begin{equation}
\xi_k^{(2)} (r) = \ofo \left(a,-1-a;1;u\right)\,,
\end{equation}
where $a=-1/2 + \sqrt{1-e^{2\lambda} k^2}/2$, $u=1-\ell^2/r^2$.
The second solution (regular at $r=0$) leads to the 
 two-point function with the discrete spectrum~\cite{Gubser}:
\begin{equation}
{\cal F}(k^2)= -\frac{N^2 k^4}{64\pi^2} \left[ \Psi \left(\frac{1}{2} +
\frac{1}{2}\sqrt{1-e^{2\lambda} k^2}\right) +  \Psi \left(\frac{1}{2} -
\frac{1}{2}\sqrt{1- e^{2\lambda} k^2}\right)\right]\,.
\label{gubser_k_n=4}
\end{equation}
We shall see that in this case the Green's function is not
monotonic with respect to the parameter $\lambda$.
The coordinate representation of (\ref{gubser_k_n=4}) is given by
\begin{equation}
G (x) = \frac{N^2}{2\pi^4 x}e^{-7\lambda} 
\sum\limits_{n=1}^{\infty} (2n+1)[n(n+1)]^{5/2}
K_1 (2\sqrt{n(n+1)}xe^{-\lambda})\,.
\label{G_gubser_discrete}
\end{equation}
The explicit expression for $G(x)$ can be 
 obtained from (\ref{G_gubser_discrete})
using Euler - Maclaurin formula:
\begin{eqnarray}
G (x) & = & \frac{ N^2 e^{-7\lambda}}{\pi^4 x} F(x,\lambda )
+
 \frac{3\sqrt{2} N^2  e^{-7\lambda}}{\pi^4 x}K_1 (2\sqrt{2}x e^{-\lambda})
 \nonumber \\
&-&
\frac{N^2 e^{-7\lambda}}{2\pi^4 x}
\sum\limits_{n=1}^{\infty}\frac{B_{2n}}{(2n)!}
f^{(2n-1)}(0)\,,
\label{G_gubser_discrete_euler}
\end{eqnarray}
where
$
f(t) = (2t+3)[(t+1)(t+2)]^{5/2} K_1 ( 2\sqrt{(t+1)(t+2)}x  e^{-\lambda})\,,
$
and the integral involved is calculated in the Appendix with the result 
\begin{eqnarray}
F(x, \lambda ) &=& \int\limits_{\sqrt{2}}^{\infty}t^6 
K_1 (2tx\lambda )dt \nonumber \\ &=&
\frac{3e^{7\lambda}}{x^7} \Biggl\{ 1 + \left(1+x^2e^{-2\lambda}\right)^2
2\sqrt{2} x e^{-\lambda} K_1 ( 2\sqrt{2} x e^{-\lambda})
\nonumber \\ 
&+& 4x^2 e^{-2\lambda}\left( 1+x^2e^{-2\lambda} +\frac{x^4e^{-4\lambda}}{3}
\right)K_0 ( 2\sqrt{2} x e^{-\lambda}) 
\nonumber \\
&-&
2\sqrt{2} x e^{-\lambda}\left[ K_0 ( 2\sqrt{2} x e^{-\lambda})
I_1 ( 2\sqrt{2} x e^{-\lambda}) +
 I_0 ( 2\sqrt{2} x e^{-\lambda})K_1 ( 2\sqrt{2} x e^{-\lambda})\right]
\Biggr\}.\nonumber
\end{eqnarray}
Representation (\ref{G_gubser_discrete_euler}) allows us to study
the region of small $x$:
\begin{equation}
G(x, \lambda ) = \frac{3N^2}{\pi^4 x^8} +
\frac{2 N^2 e^{-6\lambda}}{315\pi^4 x^2}
 + O \left( 1, \log{(xe^{-\lambda})}\right)\,.
\label{G_small_x}
\end{equation}
Note that the scale-invariant function $x^8G(x)$ is {\it not}
monotonic (see Fig. (\ref{apr113})).
 Analytically it can be seen from (\ref{G_small_x}),
 where the second term is {\it positive definite}, so that
for small $x$ the function  $x^8G(x)$ first {\it increases} before 
decreasing (exponentially) down to zero for $x\rightarrow \infty$.

Nevertheless, the canonical $c$-function {\it is} monotonic.
The  $c$-function can be obtained by slightly modifying the procedure
 described in \cite{a_note}:
\begin{equation}
c (x) = \frac{N^2x^3e^{-3\lambda}}{2} 
\sum\limits_{n=1}^{\infty} (2n+1)\sqrt{n(n+1)}K_1 (2\sqrt{n(n+1)}
xe^{-\lambda} )\,.
\label{c_gubser_discrete}
\end{equation}
Applying Euler - Maclaurin summation formula, we get
\begin{eqnarray}
c (x) & = & N^2 x^2 e^{-2\lambda}K_2 (2\sqrt{2}x e^{-\lambda}) +
 \frac{3\sqrt{2} N^2 x^3 e^{-3\lambda}}{4}K_1 (2\sqrt{2}x e^{-\lambda})
 \nonumber \\
&-&
\frac{N^2 x^3 e^{-3\lambda}}{2}
\sum\limits_{n=1}^{\infty}\frac{B_{2n}}{(2n)!}
f^{(2n-1)}(0)\,,
\label{c_gubser_discrete_euler}
\end{eqnarray}
where
$
f(t) = (2t+3)\sqrt{(t+1)(t+2)} K_1 ( 2\sqrt{(t+1)(t+2)}x  e^{-\lambda})\,.
$
The expansion for small $x$ is
\begin{equation}
c(x) = \frac{N^2}{4} - \frac{N^2 x^2 e^{-2\lambda}}{6} +
 O\left( x^4, x^4\log{x}\right)\,.
\label{c_gubser_small_x}
\end{equation}
The function $c$ is positive definite and monotonic (Fig.(\ref{apr1})).

\vskip .2in \noindent \textbf{Acknowledgments}\vskip .1in \noindent 
 M.P. is supported in part by NSF
grants no. PHY-9722083 and PHY-0070787.

\subsection*{Appendix: 
Calculation of the integral $I(c,a)=\int\limits_{a}^{\infty}x^6  
K_1 (cx )dx$  }
Volume 2 of \cite{prudnikov} (1.12.1.1, p.47)
provides an expression for an
{\it indefinite} integral, $\int x^6 
K_1 ( x )dx$, in terms of Lommel's functions which for our purposes
can be rewritten as
\begin{equation}
I(c,a) = \frac{1}{c^7}\left[ \frac{x^7}{6} K_1(x) 
\; \kofo \left( 1;4,4;\frac{x^2}{4}\right) +  \frac{x^8}{48} K_0(x) 
\; \kofo \left( 1;4,5;\frac{x^2}{4}\right)\right]^{\infty}_{ca}\,,
\label{A01}
\end{equation}
where
\begin{equation}
\kofo \left( a;b,c;z\right) = \sum\limits_{n=0}^{\infty}
 \frac{(a)_n}{(b)_n(c)_n}\frac{z^n}{n!}
\label{A1}
\end{equation}
is one of the hypergeometric functions, $(a)_n=\Gamma (n+a)/\Gamma (a)$.
It is known (see 7.14.2.98 in Volume 3 of \cite{prudnikov}) that
\begin{equation}
\kofo \left( 1;4,4;z^2\right) = \frac{9}{z^6}
\left[ 4 I_0(2z) - 4 - 4 z^2 - z^4\right]\,.
\label{A2}
\end{equation}
Integrating the series (\ref{A1}) using (\ref{A2}) we obtain
\begin{equation}
\kofo \left( 1;4,5;z^2\right) = \frac{36}{z^8}
\left[ 4 z I_1(2z) - 4z^2 - 2 z^4 - \frac{z^6}{3}\right]\,.
\label{A3}
\end{equation}
Substituting (\ref{A2}) - (\ref{A3}) into (\ref{A01}) we get
\begin{eqnarray}
\int\limits_{a}^{\infty}x^6 
K_1 (cx)dx &=&
\frac{384}{c^7}\Biggl\{ 1 + \left( 1+\frac{a^2c^2}{8}\right)^2 ac K_1 (ac) +
\frac{a^2c^2}{2} \left( 1 + \frac{a^2c^2}{8} +\frac{a^4c^4}{192}\right)
K_0(ac)\nonumber \\ & -& ac \left[ K_0(ac)I_1(ac)+I_0(ac)K_1(ac)\right]\Biggr\}\,.
\label{res}
\end{eqnarray}
For $a=0$ (\ref{res}) reduces to the well-known result,
$I(c,0)=384/c^7$ \cite{prudnikov}.

\end{document}